# MTTBA- A Key Contributor for Sustainable Energy Consumption Time and Space Utility for Highly Secured Crypto Transactions in Blockchain Technology

M. Gracy[1], B. Rebecca Jeyavadhanam[2] [*]

[1] SRM Institute of Science and Technology, Kattankulathur, India

[2] SRM Institute of Science and Technology, Kattankulathur, India

[*]corresponding author. Email address : rebeccaram2022@gmail.com

**ABSTRACT**

A Merkle tree is an information construction that is used in Blockchain to verify data or transactions in a large content pool in a safe manner. The role of the Merkle tree is crucial in Bitcoin and other cryptocurrencies in a Blockchain network. In this paper, we propose a bright and enhanced verification method, Merkle Trim Tree-based Blockchain Authentication (MTTBA) for the hash node traversal to reach the Merkle Root in a minimum time. MTTBA is a unique mechanism for verifying the Merkle Tree's accumulated transactions specifically for an odd number of transactions. The future impact of cryptocurrency is going to be massive and MTTBA proves its efficacy in transaction speed and eliminating node duplication. Our method enables any block to validate transactions' full availability without duplicating hash nodes. Performance has been evaluated in different parameters and the results show marked improvement in throughput(1680ms), processing time(29700kbps), memory usage(140MB), and security(99.30%). The energy consumption factor is crucial in the scenario, and MTTBA has achieved the lowest of 240 joules.

## 1 Introduction

The internet has become an integral aspect of data transmission in our daily lives as a result of the improved development of Information Technologies. Blockchain Technology was first presented as a public, decentralized, and trust-less digital currency ledger [1] and it has gained prevalent acceptance in many fields. Blockchain [2] (e.g., Bitcoin and Ethereum) chronologically keep track of transactions, grouped into a chain of blocks. Network nodes add to the ledger by creating and adding new blocks, starting with the genesis block. Full nodes that download the whole block tree validate the transactions in the received blocks. However, a blockchain must include light nodes [3], which may just be interested in confirming a few specific transactions, for better scalability. In the original Bitcoin protocol [4], which is a public database of financial transactions, the blockchain is used to keep track of coins [5]. The ledger network is also known as a decentralized peer-to-peer network [2]. The characteristics of bilinear mapping can be used to achieve data integrity in the form of blockchain transactions [6].

In this paper, we proposed a new optimized Merkle tree structure called MTTBA, to do the verification process faster and easier with no duplication of nodes. This idea makes transaction verification easier especially when the block consists of an odd number of transactions. To solve the space occupied by the duplicate nodes in the traditional Merkle tree method, the proposed technique, MTTBA handles the problem in a novel way and helps in finding the tampered node with a quick traversal of nodes. The traditional Merkle tree uses node duplication when the block is accumulated with an odd number of transactions. It needs extra memory space to accommodate the duplicate nodes. Usharani et al suggested an idea [7], the Modified Merkle Tree data structure, which is used to design a system that handles data authentication, consistency verification, and data synchronization. According to our design, the number of hash nodes appending to reach the root is reduced and storage space is considerably reduced. The proposed MTTBA has high throughput with low energy consumption and has no compromise in terms of energy consumption, processing time utilization, and security. The authentication process is inevitably fast [8]. The results and analysis section will deliberately prove that this MTTBA variant is superior in all parameters.

Further sections can be divided as follows: Section 2 gives the background of the Blockchain, Merkle tree and Hashing. Related work presented in Section 3, while Section 4 explains about the proposed method. Implementation of the proposed work is presented in Section 5. Finally, the Conclusion and Future directions are in Section 6.

## 2 Background

### 2.1 *Blockchain*

Cryptography, Mathematics, Consensus Algorithms, and Economic Models are among the technologies that make up Blockchain Technology [9]. It is a safe distributed ledger (dataset) that records all transactions as blocks. The blockchain uses peer-to-peer networks and consensus procedures to solve the problem of distributed data synchronisation, eliminating the need for a trusted centralized authority. A SHA256 cryptographic hash algorithm on the block header can be used to identify each block[10]. Bitcoin is one of the most well-known blockchain-based applications. Each transaction includes information on the sender and receiver and the number of coins to be exchanged. Once confirmed by peers, a group of such transactions forms a new block[11]. Fuad Shamieh et al. [12] envisioned a heuristic for selecting transactions from miners' transaction pools to form blocks with a desired operational value to achieve predefined production outputs over blockchain-based networks. The primary data is a list of transactions [13], whereas the header is a hash of the previous and current blocks, Merkle root, timestamp, nonce, and other information. Blockchains are used in many scenarios like healthcare, IoT[14], and many more. Figure 1 gives the structure of the Block. The basic features of Blockchain [15] [16]are transparency, decentralization and immutability.

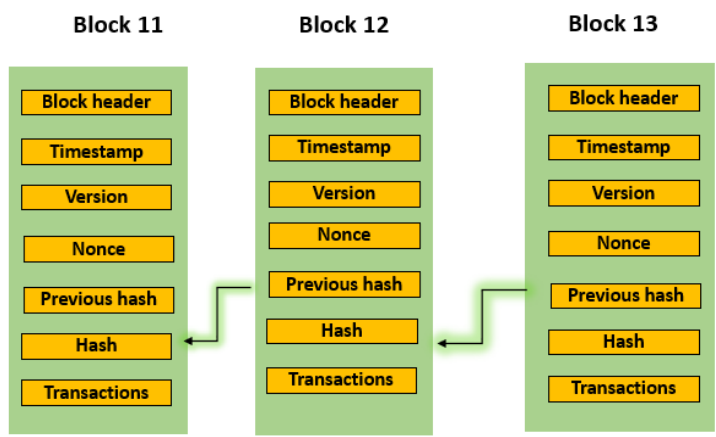

Figure 1. Blockchain Structure

## 2.2 The Merkle Tree

The Merkle trees are information constructions that are used to verify data in a big content pool in a secure manner. It is an awesome concept and acclaim for coming up with a nice idea goes to Ralph Merkle [4]. Merkle roots [17] are vital in the computation required to keep cryptocurrencies such as bitcoin and ether operational. In their paper, Haojun Liu et al. [18] explain systematical insights into Merkle trees in terms of their concepts, attributes, benefits, and applications.. Merkle trees also provide a way to authenticate [19][20]. The traditional Merkle tree construction is shown in Figure 2.

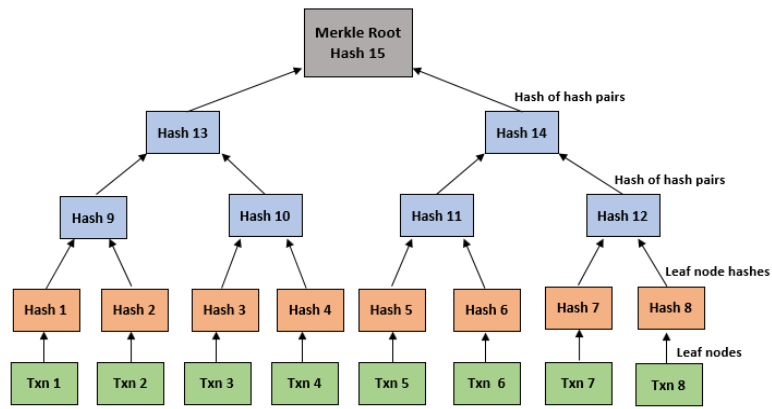

Figure 2. Traditional Merkle tree structure

All blocks are hashed using cryptographic string hash methods, such as SHA256 [21]. The data is stored at the leaf nodes of a Merkle Tree, which is a binary tree[22][23]. The Merkle hash tree technique can be used in securing smart grid communication with smart meters having computation-constrained resources [24] and enhancing security with timestamps[25]. Figure 3 shows the clear architecture of the Merkle root, which forms from the group of transactions in Block 3.

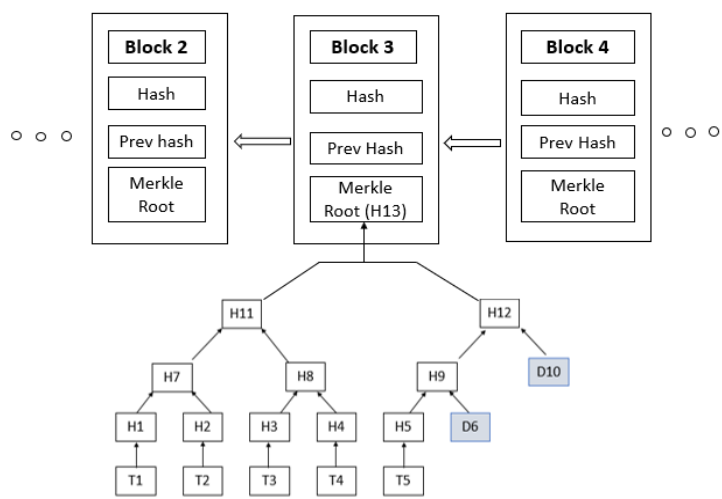

Figure 3. Merkle root architecture

Figure 4 shows five transactions and the Merkle root is formed from the five transactions. Green-colored hash nodes represent the verification of the transaction T3, which is in orange color. The hash nodes involved in verifying T3 = H4, H7, H12

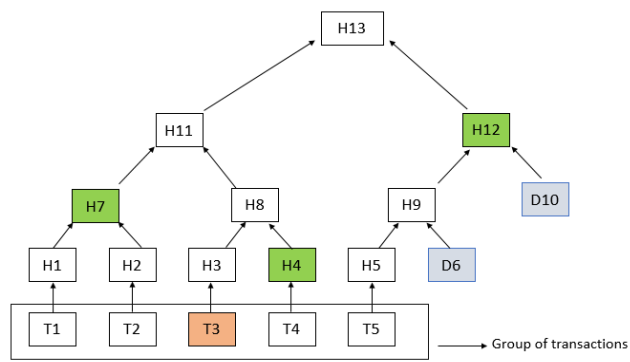

Figure 4. Transaction verification in Merkle tree

*2.3 Hashing*

The hashing process is critical to the data integrity of blockchain. Hashing is a one-way fixed-size visualisation that transforms any size of the file into a distinctive, collision-restricted code. Ashok Kumar Yadav [26] uses random private keys to implement ECC and RSA public generation, encryption, and decryption. Yoon-Su Jeong[27] demonstrated an optimised hash computation method that enables for stratified multiple processors on an n-bit blockchain. Monika et al. [28] propose that hash functions be studied and that different cryptography algorithms be compared utilizing blockchain technology.

3 **Related Work**

In this section, we present a detailed survey on related papers with the Merkle Tree and transactions in the blockchain. Dong et al found a new way in the form of Merkle tree-based operational online authentication[29]which tells a simple modification that involves the addition of random inputs with two tangibly separated entities, prover and approver. Merkle Quad-Tree technique [30] is a method that displays steady advantages with various levels of modifications and consumes about 1% of the time of the traditional Merkle Tree method. Similar to Lamports' authentication approach, Yuji et al proposed a method [31] in which the authentication act is carried out by releasing the digest values associated with nodes that are higher in the hash chain than the digests of the leaf node. Lum Ramabaja et al. [32] proposed a standard sparse Merkle multi-proof that involves storing an index for every non-leaf hash in the multi-proof. To replace the Merkle tree, which has network delay, Patgiri et al. developed an alternative model called Hex-Bloom [33]. Jia Kan et al [34] present MTFS, a blockchain-based solution for private file storage. Markus et al [35] propose an efficient algorithm for traversing Merkle trees, as well as a technique for producing a classification of leaves and their associated authentication paths. Using Merkle tree authentication, author Dong et al. [36] presented a novel technique to safeguard outsourcing data. Wang et al. [37] propose a revised Merkle tree structure for efficacious payment transactions in blockchain-based IIoT systems. Ceaser Castellon et al.[38] proposed an idea that employs an energy-saving algorithmic engineering technique for Merkle Tree root calculations, a key component of blockchain simulations, to maintain the predicted protection while sacrificing less system availability. Rasmus et al. proposed a sparse Merkle tree definition that allows efficient space-time exchange for different caching strategies when using SHA-512/256. For to reduce the bandwidth required for Merkle Trees, John Kuszmaul [39] proposed a new data structure called Verkle Trees, but this comes at the cost of increasing the computational power. Bruschi et al.[40] proposed a system based on a Merkle tree representation to yield compact justification of the content of web documents. Bayardo et al. [41] proposed using Merkle trees to support 200 and 404 response authentication while requiring only a single cryptographic hash of trusted data per storage system. Table 1 details the investigated five existing methods, including their methodology, benefits, and drawbacks. Section 5 contains the detailed experimental results of the same.

| S.No | Author | Title | Methodology | Benefits | Drawbacks |
|---|---|---|---|---|---|
| 1. | Dongyoung Koo et al. 2018 | MTDALR[29] | Inserting ancillary random sources into the integrity verification proof on the prover side. | Maintains steady reliability without being harmed by continual data leakage caused by authentication process repetitions. | More energy consumption and more memory space are needed. |
| 2. | Pavol Zajac 2021 | EKAMT[42] | A server-authenticated key that is appropriate for TLS-like handshake protocols. | Provides efficient protocols suitable for Internet of Things application | Use in constrained devices limited due to the operating time and memory requirements. |
| 3. | Yi-Cheng Chen et al. 2019 | IASMTM[43] | Tampered region detection on the tampered image using peak signal-to-noise ratio value | Substantiates the reliability of the image and repairs the damaged area in the image. | Takes more time for authentication, Less Throughput time, and more end-to-end delay. |
| 4. | Teasung Kim et al. 2020 | SELCOM[44] | SELCOM is used to solve the storage problem for blockchain nodes with limited resources. | It allows each node to choose to maintain blocks through the proposed checkpoint chain. | Security is a concern and processing time is not high nor low. |
| 5. | Mingchao Yu et al. 2019 | CMT[45] | Using a peeling-decoding technique, a succinct proof for data availability attack on any layer is possible. | Any node can validate the entire availability of every data block generated by the system. | It has an average throughput time and needs more memory space |

Table 1. Existing papers methodology and its advantages and limitations for Experimental Analysis

## 4 Proposed Method

### 4.1 Preliminaries

We denote $t_n$ as the total number of nodes for the scheme. Let n={1,2,3,..} be the number of nodes. For any n≥1ϵ N, [N] denotes a set of even number of transactions. N= $\{2^1, 2^2, 2^3, 2^4, ..2^n\}$. for any n≥1ϵ N-1, [N-1] denotes a set of the odd number of transactions. N=$2^n$ and N-1 transactions are said to be the odd number of transactions.

### 4.2 Construction of the Proposed Scheme

Traditional Merkle trees are time-consuming data structures that waste a significant amount of computational resources. Figure 5 depicts the MTTBA construction, where node duplication is commletely removed. After successfully downloading a data block using Merkle root, it permits verification of the data block's validity and integrity and takes a few hash values and does not impose the Merkle tree on its whole. A block can accommodate 1MB of transactions, if the number of transactions is even in number, the construction of the Merkle tree is done traditionally. If the block holds an odd number of transactions, then our proposed scheme MTTBA will be a handy one. As the Blockchain has a high level of data integrity, transactions accommodated in a block cannot be manipulated by any third person. The construction steps of the MTTBA for the block are as follows:

1) Set index (1) for the first node n1 with hash 1 in the Merkle Trim Tree for a block.
2) Append left and right nodes consecutively leaving the first block, index (1) all alone for any $t_n$ transactions,

    for example, if the block has 2810 transactions, the hash nodes will keep on appending in pairs until it reaches 2810 leaving the first hash node aside.
3) From 1 to $t_n$, apply the hash function to each of the divided nodes.
4) After that, build the top level of the Merkle Tree until the Merkle Root is reached.

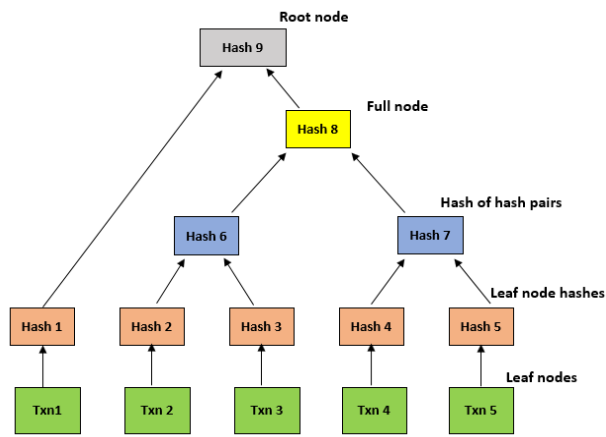

Figure 5. Construction of Merkle Trim Tree-Based Blockchain Authentication Verification Method

Furthermore, the traditional Merkle tree has significant time complexity and the Merkle tree server takes a large amount of memory to keep all of the hash values; nevertheless, a user does not need to save the complete Merkle tree. In contrast to traditional Merkle Tree construction and various proposed ideas for verification and validation of transactions, we proposed a method called Merkle Trim Tree-based blockchain authentication, which is constructed using a combination of the Merkle Tree's fundamental principle and the Merkle Trim Tree partition methods.

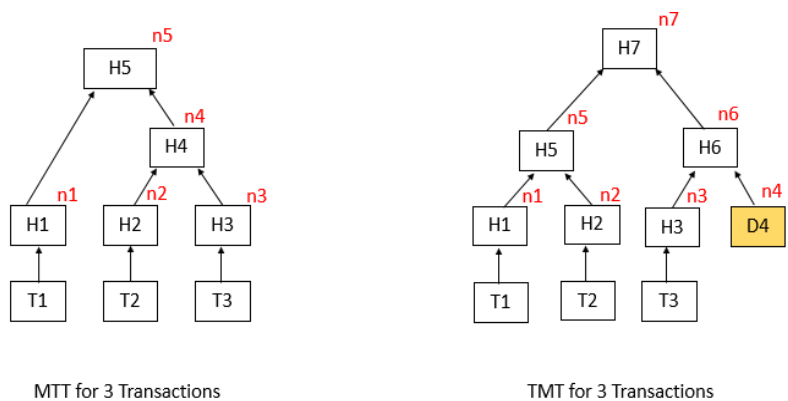

Figure 6a. The comparison of transactions in the traditional Merkle tree (TMT) and Merkle Trim tree (MTT) to form the root for three transactions.

The comparison of transactions in the traditional Merkle tree (TMT) and Merkle Trim tree (MTT) to form the root is shown in Figures 6a and 6b for three and five transactions respectively. The difference in the nodes has been shown clearly in 'n'.

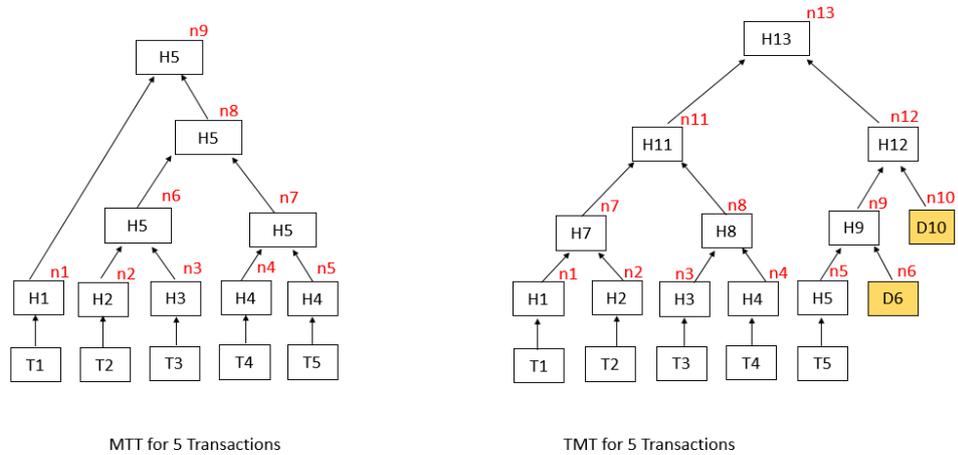

Figure 6b. The comparison of transactions in the traditional Merkle tree (TMT) and Merkle Trim tree (MTT) to form the root for five transactions.

Merkle Trim Tree (MTT) stands tall as it can eliminate duplication as well as the number of nodes it creates for forming the tree transactions. When the Merkle Trim Tree is compared with Traditional Merkle Tree, the duplication of nodes is eliminated and the number of nodes has been reduced.

The number of nodes reduced can be shown in an equation

$$\text{Number of nodes } t_n = n + (n-1)$$

$$t_n = 2n - 1$$

Our proposed method of MTTBA helps in transaction verification and its Authentication. The primary contribution is in eliminating node duplication. The objective of the Merkle Trim Tree is to take out the duplication nodes when a block accommodates an odd number of transactions. When the nodes are fewer, traversal becomes fast. The node duplication happens when the block has an odd number of transactions. MTTBA has incredible security when nodes are verified. The Merkle Trim Tree is structured as, leaving the first node and pairing starts from the second node onwards. If the final node in the particular block is left with none to pair, the first node will pair with the left-out node. This scenario happens when the transactions accumulated in the block were odd in number. The Merkle Trim Tree-based blockchain authentication is an efficient data structure for the verification of nodes and checking the integrity of data when transactions are odd in number. Our implementation results show security is high and the packet delivery rate is glaringly good with very less collisions. The overview of the proposed system with the neat setup and the different cases we are going to consider the deal with the methodology.

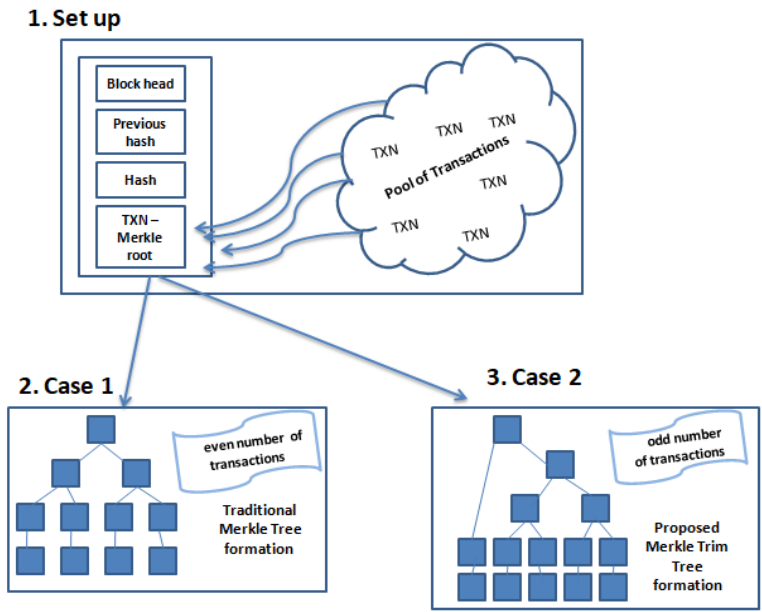

Figure 7. Overview of the proposed scheme

*Setup*: Blockchain consists of a series of blocks in chronological order. Every block will have components like block header, nonce, timestamp, hash, previous hash, and transactions. Consensus mechanisms are used to validate a blockchain and add new blocks. Transactions will be selected to join the block from the pool of transactions. This has been shown in Figure 7. A block can hold 1MB of transactions and after that new block will be appended and transactions continue to fill. The accommodated transactions will be accessed with a single hash called Merkle root. The transactions can be odd or even in number. Our proposed work gives the methodology for verification of the odd number of transactions. The verification of even the number of the transaction remains the same as the traditional one.

*Case 1*: Even the number of transactions for $N=2^n$ transactions, will be an even number of transactions, and no duplication is needed till reaches the root.     n=1, 2, 3… nodes $N = 2^1, 2^2, 2^3,…2^n$ . [ 2,4,8,16,32,64,128,…] these number of transactions no need duplication.

*Case 2*: Odd number of transactions. Needs duplication in the first level for appending.
$N=2^n$ and N-1 transactions are an odd number of transactions.  if n=4, then $2^4 = 16$ transactions and in case 2, it is N-1 and here 15 transactions. [1,3,5,7,9,11,…] these number of transactions need duplication in the first level itself to find the pair to concatenate.

**4.4.1** *Algorithm*

---

Set up the scenario for transaction verification.
  N = set of even numbers, N-1 = set of odd numbers
   $t_n$ is the total number of transactions,
   H is the hash of the node.

---

If $t_n =N$, then the block contains an even number of transactions
For any n≥1 and 1 ϵ N, n++
Append $t_1(H_1)$ and $t_2(H_2)$ to form $H_5$. $H_5$ = hash of $n_1$ and hash of $n_2$ ($H_1+H_2$)
Append $t_3(H_3)$ and $t_4(H_4)$ to form $H_6$. $H_6$ = hash of $n_3$ and hash of $n_4$ ($H_3+H_4$)
$H_7 = H_5+H_6$, $H_7$ is the root node or $H_n$ is the Merkle root

The number of nodes for the set of even numbers is $t_n=2n-1$

If $t_n = N-1$, then the block contains an odd number of transactions
$t_1(H_1)$ is left idle for initial appending
For any $n \geq 1$ and $1 \in N-1$, n++
Append $t_2(H_2)$ and $t_3(H_3)$ to form $H_6$. $H_6$ = hash of $n_2$ and hash of $n_3$ $(H_2+H_3)$
Append $t_4(H_4)$ and $t_5(H_5)$ to form $H_7$. $H_7$ = hash of $n_4$ and hash of $n_5$ $(H_4+H_5)$
$H_8 = H_1+H_7$, $H_8$ is the root node or $H_n$ is the Merkle root

The number of nodes for the set of odd numbers is $t_n=2n-1$

## 5. Implementation of MTTBA

All experiments were run on a single machine running Windows 10 with a 2.6 GHz CPU (Intel i5-3320M) and 16.0GB RAM (2601MHz 2x8 GB). Using the Hyper Caliper, a crypt analytic tool, each algorithm was built as a Visual Studio 16.0 program. In addition, when the Merkle Trim tree was built, the data was separated into 256-byte blocks to allow for consistent comparison. Each parameter was repeated numerous times in the same environment to produce the optimized report comparing the proposed approach with the existing method, and then the parameters were calculated and reported. The computation time for each experiment was measured based on CPU time. The performance of each algorithm for varying data sizes is analyzed and compared with various parameters like throughput, end-to-end delay, average processing time, average authentication time, average energy, packet delivery ratio, security, and memory utilization.

### 5.2.1 *Throughput*

MTTBA's throughput time has proven to be consistent, thus it was included in the experiment as a baseline metric of efficiency. Figure 8 shows the graphical representations to prove the proposed MTTBA outperforms in throughput when compared with other existing methods.

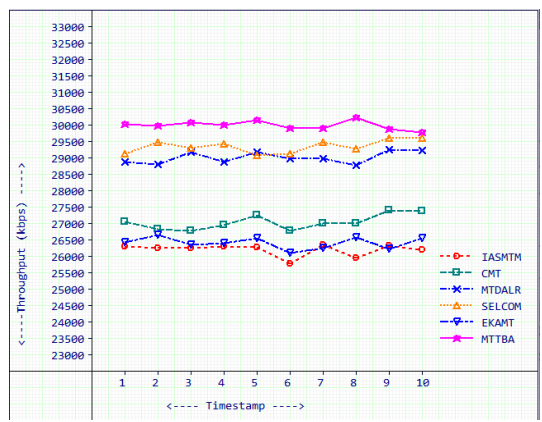

Figure 8. Throughput time

### 5.2.2 *Average Processing Time*

Processing time for a job is very important and this will define a genuine work methodology. In 1750 milliseconds, MTTBA performs the given task. This can be verified in various timestamps. Figure 9 gives a detailed representation of the graph.

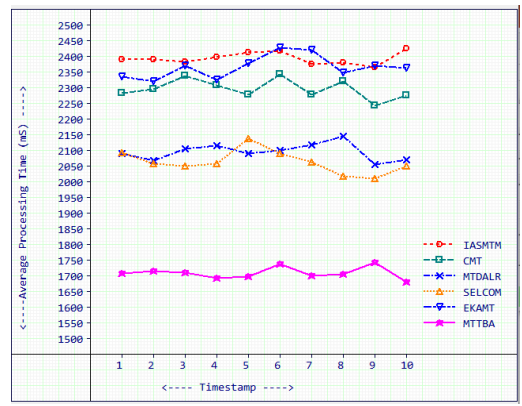

Figure 9. Average Processing Time

### 5.2.3 *Average Energy*

In Blockchain, energy consumption [46] [47] is an important factor to be considered and different methods consume different joules of energy when the process runs in different timeframes. Figure 10 shows clear evidence of low energy consumption by MTTBA with the lowest of 240 joules.

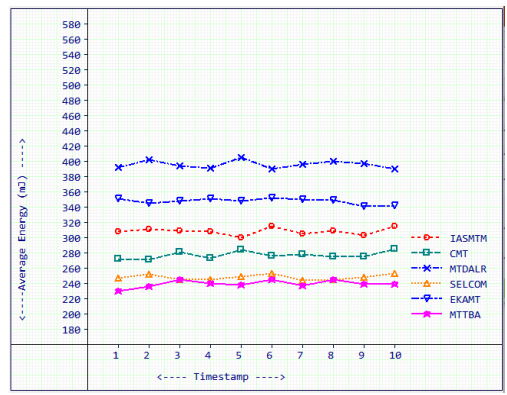

Figure 10. Average Energy Consumption

### 5.2.4 *Average Authentication Time*

The average authentication time of the proposed MTTBA is less than 100 milliseconds when compared with IASMTM having maximum authentication time and MTDALR having medium authentication time. The comparisons have been shown in Figure 11.

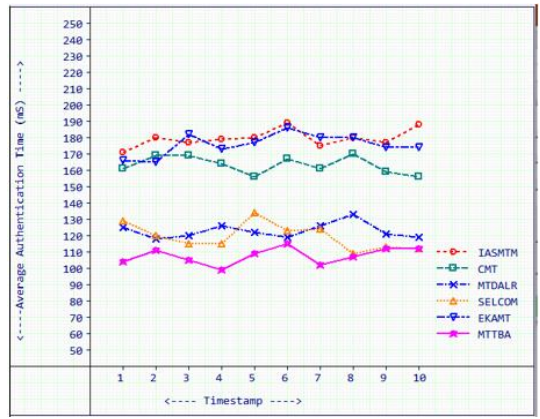

Figure 11. Average Authentication Time

### 5.2.5 *Security*

Security plays a pivotal role in any application or tool, specifically in the decentralized form of systems. Blockchain is a decentralized application and the security of data plays a major role. Figure 12 shows the ability of MTTBA, which stands tall in security. MTTBA never compensates for security with any metrics.

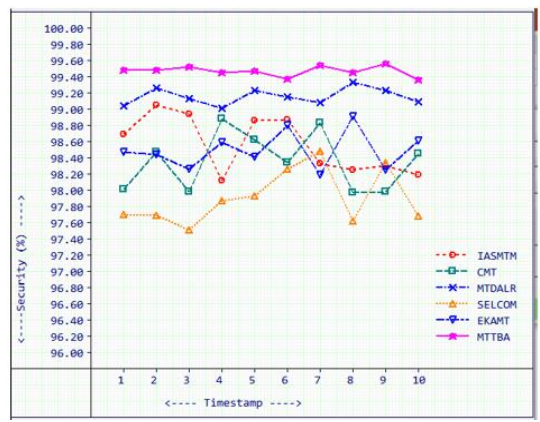

Figure 12. Security

### 5.2.6 *Memory*

Memory is the talk of argument whenever a new application or tool or method is developed. Blockchain technology never hesitates this parameter. Blockchain's main issue is its scalability since every data inside the block will have a hash value and this will be duplicated as a ledger. Chances are there for more memory consumption and relevant experiment in Figure 13, where our proposed MTTBA consumes less memory when compared with our existing methods. This can be easily achieved by MTTBA as it completely obliterates the duplication nodes in the traditional Merkle Tree.

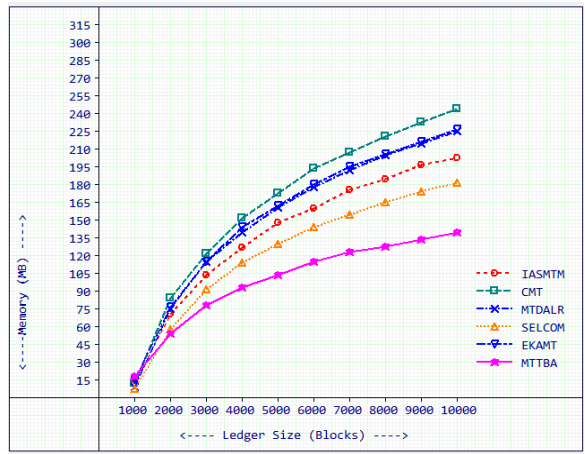

Figure 13. Memory Consumption

The convenient way of sending data in the network is via packets, but the primary problem that arises during packet delivery is data collision. If 100 or more packets are sent at a time, then there is a chance of collision. Our proposed MTTBA variant's Packet delivery ratio is very high. If the packet delivery ratio is high, then the latency is obliviously low. Because throughput in KBPS is directly proportional to the packet delivery ratio.

**6 Conclusion and Future Work**

The results of the studies show that hash duplication has been taken out with the help of the Merkle Trim Tree construction. The Merkle Trim Tree structure, especially when an odd number of transactions is present in a block, helps to reduce duplication by appending with the initial transaction. Energy consumption is always in check and MTTBA with its unique style can able to achieve its goal of transaction verification more competitively. The data appending method proposed in MTTBA results in limited and constant size and appears to be the source of the significant improvement. Memory usage and Security help MTTBA to dominate the entire experimentation process. The Merkle Trim Tree aiding approach, on the other hand, is identical to the different degrees of alterations and takes around 1% of the time of the old way. In future work, we will merge Merkle Trim Tree with the Bloom-filter algorithm to create an advanced Merkle Trim Tree that can arrive with a high level of data integrity as its main focus.

**Author Contributions**: Project administration, R.J.; Supervision, M.G.; Writing-original draft, M.G.; Writing-review and editing, M.G. and R.J. All authors have read and agreed to the published version of the manuscript.

 **Funding**: This research received no external funding.

**Institutional Review Board Statement**: Not applicable.

**Informed Consent Statement**: Not applicable.

**Data Availability Statement**: Not applicable.

**Conflicts of Interest**: The authors declare no conflict of interest